\documentstyle[prl,aps,twocolumn,psfig,floats]{revtex}

\begin{document}

\draft

\title{Time-dependent control of ultracold atoms 
and Bose-Einstein condensates in magnetic traps} 

\author{N. V. Vitanov$^1$ and K.-A. Suominen$^{1,2}$}

\address{(1) Helsinki Institute of Physics, PL 9,
FIN-00014 Helsingin yliopisto, Finland}

\address{(2) Theoretical Physics Division, Department of Physics, University
of Helsinki, PL 9, FIN-00014 Helsingin yliopisto, Finland}


\maketitle

\begin{abstract}
With radiofrequency fields one can control ultracold atoms in
magnetic traps. These fields couple the atomic spin states, and are used
in evaporative cooling which can lead to Bose-Einstein condensation in
the atom cloud. Also, they can be used in controlled release of the
condensate from the trap, thus providing output couplers for atom lasers. In
this paper we show that the time-dependent multistate models for these
processes have exact solutions which are polynomials of the solutions of the 
corresponding two-state models. This allows simple, in many cases analytic 
descriptions for the time-dependent control of the magnetically trapped 
ultracold atoms.
\end{abstract}

\narrowtext

\vspace{0mm} 

\begin{figure}[htb]
\vspace*{-1cm}
\centerline{\psfig{width=80mm,file=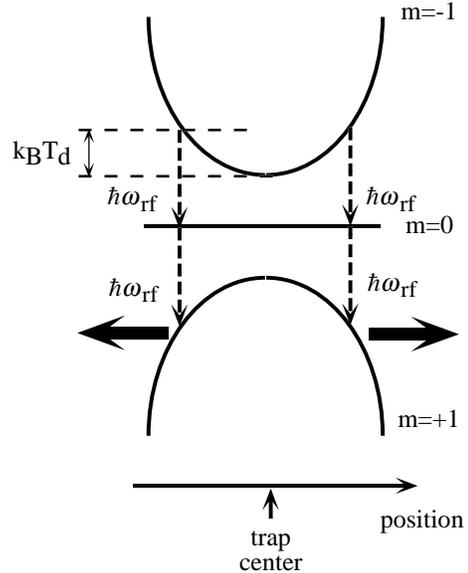}}
\vspace*{-25mm}
\caption[f1]{Magnetic trapping of neutral atoms. We show the atomic
potentials for the $f=1$ trapping system, and how the radiofrequency field
(frequency $\omega_{\rm rf}$) can be used to control the trap depth $k_B T_d$.
The resonant transition $m=-1\rightarrow m=0
\rightarrow m=+1$ moves the atoms with kinetic energy larger than $k_B
T_d$ from the trapping state to a strongly nontrapping state. 
\label{yksi}}
\end{figure}

Neutral atoms possessing hyperfine structure can be trapped in spatially
inhomogeneous magnetic fields, if they are in the appropriate spin state. As
the magnetic field imposes spin-dependent Zeeman shifts on the atomic energy
levels, a spatially changing magnetic field maps into an external potential
felt by the atom (see Fig.~\ref{yksi}). The magnetic traps are, however, very
shallow, so only ultracold atoms can be trapped. For alkali atoms the proper
temperatures have been obtained via precooling with laser light. Once
the atoms are trapped, one can decrease the trap depth, which allows the hot
atoms to escape, and those left behind thermalize via collisions into a lower
temperature---this is called evaporative cooling~\cite{evap}. By combining
magnetic traps with evaporative cooling one can now reach the densities and
temperatures where Bose-Einstein condensation takes place~\cite{BEC}. Then the
atoms form a coherent superposition, which can be released from the
trap~\cite{MITcoupler}. As the escaping atoms maintain their
coherence~\cite{MITalaser}, the experiment is a prototype for an
atom laser, i.e., production of coherent, propagating packets of matter waves. 

\begin{figure}[htb]
\centerline{\psfig{width=100mm,file=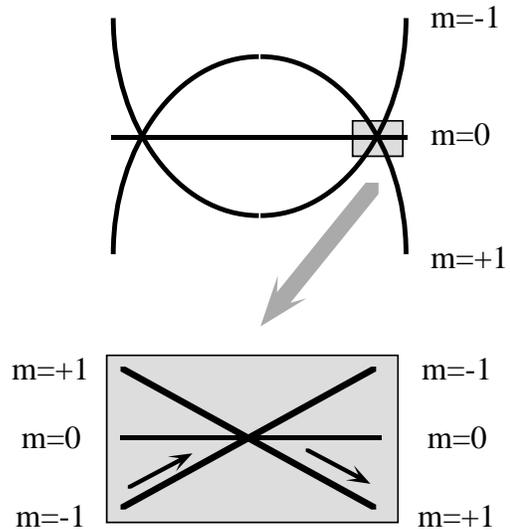}}
\vspace*{-6cm}
\caption[f2]{The curve crossing description of trap dynamics. 
The potentials are obtained by shifting the
atomic states by multiples of rf photon energies. The region near
the resonance point can be modelled with a bowtie crossing as
shown in the enlargement of the grey area. Assuming that an atom traverses the
crossing with some constant velocity $v_c$, we can map the position-dependent
crossing into a time-dependent one, which in the special case of two states
corresponds to the widely used and well-known Landau-Zener
model. In the optimal, i.e., adiabatic case the hot atoms follow the route
marked by arrows out of the trap.
\label{kaksi}}
\end{figure}

Evaporative cooling requires effective and precise control of the trap depth.
This can be achieved by coupling the spin states (labeled with the magnetic 
quantum number $m$) with a radiofrequency (rf) field~\cite{evap,rf,Davis}. The 
coupling introduces transitions between the neighbouring spin states ($\Delta
m=\pm 1$). As demonstrated in Fig.~\ref{yksi}, the frequency $\omega_{\rm rf}$ 
of the field controls the depth of the trap. If we want to estimate the 
efficiency of evaporative cooling, we can transform the rf resonances into 
curve crossings, and describe the dynamics of the atoms at the edge of the trap 
with an appropriate time-dependent curve crossing model~\cite{evap}, 
see Fig.~\ref{kaksi}. Typically one compares the trap oscillation period times 
the spin-change probability at the resonance point to the other time scales of 
the trapping and cooling process, including collisional loss rates~\cite{evap}.

However, as one usually operates in the region where Zeeman shifts are linear,
the rf field couples sequentially all spin states, instead of just selecting a
certain pair of states. In case of only two spin states we can use
the standard Landau-Zener model in estimates of the efficiency of evaporative
cooling. In practice, however, one has $2f+1$ states, where $f$ is
the hyperfine quantum number of the atomic state used for trapping. So far
condensates  have been realised for $f=1$ and $f=2$, but experiments e.g.~with
cesium involve states with $f=3$ and $f=4$, so in order to achieve efficient 
evaporative cooling one needs several sequential rf-induced transitions. 
Clearly the use of the two-state Landau-Zener model can be questioned in these 
multistate cases.

Once the Bose-Einstein condensation has been achieved one can release the 
condensate just by switching the magnetic field off. This technique, however, 
does not allow much control over the release. Moreover, it always involves all 
the atoms. With rf fields one can transform parts of the condensate into 
untrapped states, in which they are typically accelerated away---this has been
demonstrated experimentally~\cite{MITcoupler}, as well as the fact that the
released atoms are in a coherent superposition~\cite{MITalaser}, thus 
justifying the term ``atom laser". This output coupling process can be achieved
either by using rf pulses which are resonant at the trap center, or by using 
chirped rf fields. Both correspond to a multistate Hamiltonian, where either 
the diagonal terms (chirping) or the off-diagonal terms (rf pulses) have 
explicit time dependence. In Ref.~\cite{MITcoupler} the output coupling process
was demonstrated experimentally for the sodium $f=1$ situation, and the 
transition probabilities were in good agreement with the predictions of
time-dependent three-state models. 

Both evaporative cooling and output coupling demonstrate the need to
have analytic solutions for the time-dependent multistate models of the
rf-induced dynamics. Although these models can be easily solved numerically,
they are often used only as a part of a bigger theoretical description, in
which the dependence of the solutions on the parameters such as the frequency
and intensity of the rf field, or chirp parameters and shapes of pulse
envelopes are required. Instead of looking into the known effects of rf
fields, we can consider the effects first, and then look how we need to tailor 
the rf field in order to achieve what we want---in this approach the analytic 
models show clearly their supremacy. Furthermore, as we show in this Letter, the
description of the rf-induced multistate processes is closely connected to the
two-state processes, which means that the wealth of knowledge on two-state
models that has been accumulated in the past~\cite{Shore,GS} can be 
applied---and tested---with Bose-Einstein condensates.

First we need to derive the Hamiltonian describing the rf-induced processes.
The field ${\bf B}=B_0\cos(\omega_{\rm rf}t){\bf e}_{\rm rf}$ couples to the 
atomic magnetic moment $\boldmath\mu$, i.e., $H_{\rm int}= -\mbox{\boldmath
$\mu$}\cdot {\bf B}$. The matrix
elements of this coupling between the magnetic states of the same hyperfine
manifold are nonzero only if $\Delta m=\pm 1$. Furthermore, using the angular
momentum algebra we see that the couplings between neighbouring states have
the form~\cite{evap,RoseZare}
\begin{equation}
   H_{m,m+1} = H_{m+1,m}= \sqrt{(f-m)(f+m+1)}\hbar\Omega,
\end{equation}  
where $\Omega$ is the Rabi frequency quantifying the coupling. Here we
have applied the rotating wave approximation and eliminated the field terms
oscillating with frequency $\omega_{\rm rf}$. This leads to the
curve crossing picture of the atomic potentials~\cite{GS}. 

In the regime of the linear
Zeeman effect the energy difference between two neighbouring $m$ states is
$E_Z(R)$, which is independent of $m$ but shares the $R$-dependence of the 
trapping field. Here $R$ is the distance from the trap center. Then
\begin{equation}
   H_{mm}(R) = m\varepsilon(\hbar\omega_{\rm rf}-|E_Z(R)|) \equiv 
               m\varepsilon\hbar\Delta(R).
\end{equation}
Thus $\Delta(R)$ is the local detuning of the rf field. Since the trapping state
can be either $m=-f$ or $m=f$, depending on the particular atomic system, we
need the factor $\varepsilon$, which is $+1$ for $m=-f$ trapping state, and 
$-1$ for $m=f$ trapping state. 

In order to model the multistate dynamics we seek the solution of the 
Schr\"odinger equation for an $N$-state system 
($\hbar\equiv 1$):
\begin{equation}
   \label{SEqs}
   i\frac d{dt}{\bf c}={\bf Hc},
\end{equation}
where ${\bf c}=(c_1,c_2,\ldots ,c_N)^T$ is the state vector containing the
amplitudes for each spin state. For practical reasons we label the states with
$n=1,2,\ldots N$, instead of using the $m$ labels $(m=n-1-f=-f,-f+1,\ldots
,f;N=2f+1)$. The matrix elements of the model Hamiltonian are given by
\begin{eqnarray}
   \begin{array}{lcl}
      H_{nn}(t)&=& m\varepsilon\Delta (t), \\ 
      H_{n,n+1}(t)&=&H_{n+1,n}(t)=\sqrt{n(N-n)}\Omega (t), \\ 
      H_{nk}(t)&=&0,\qquad \left(\left| n-k\right| \geq 2\right).
   \end{array}
   \label{elements}
\end{eqnarray}
Note that for a moving atom the $R$-dependence in $\Delta$ can be mapped into
time-dependence using a classical trajectory. 

We assume that initially the system is in the trapping state, which corresponds
to either $n=1$ or $n=N$, depending on $\varepsilon$. However, due to the 
symmetry of the model we solve it for the case
$\varepsilon=+1$, i.e., start with the initial conditions
\begin{equation}
   c_1(-\infty) = 1,\qquad c_{n>1}(-\infty)=0. \label{cond}
\end{equation}
Then the case $\varepsilon=-1$ is obtained by reversing the state labelling and
the sign of $\Delta$.

The model~(\ref{elements}) is a generalization of the Cook-Shore model, where 
$\Delta$ and $\Omega$ are time-independent~\cite{Cook}. We show that the 
solution for the $N$-state model with the initial conditions~(\ref{cond}) can be
expressed using the solution $(a_1,a_2)$ of the two-state equations
\begin{equation}
   \label{2SEqs}  i\frac d{dt}\left[ 
   \begin{array}{c}
      a_1 \\  a_2
   \end{array} \right] =\left[ 
   \begin{array}{cc}
      -\frac{1}{2}\Delta  & \Omega  \\ \Omega  
      & \frac{1}{2}\Delta 
   \end{array} \right] \left[ 
   \begin{array}{c}
      a_1 \\ a_2
   \end{array} \right].
\end{equation}
Moreover, our derivation is considerably simpler and more straightforward than 
the one in Ref.~\cite{Cook}, where the Hamiltonian is diagonalised by means of 
rotation matrices using the underlying SU(2) symmetry of the model.

We begin with $N=3$. The Schr\"odinger equation is
\begin{equation}
   \label{3SEqs}i\frac d{dt}{\bf c}=\left[ 
   \begin{array}{ccc}
      -\Delta  & \Omega \sqrt{2} & 0 \\ \Omega \sqrt{2} & 0 & \Omega 
      \sqrt{2} \\ 0 & \Omega \sqrt{2} & \Delta 
   \end{array} \right] {\bf c}.
\end{equation}
We make the ansatz $c_1=\lambda _1a_1^2, c_2=\lambda _2a_1a_2, 
c_3=\lambda_3a_2^2$, substitute it in Eq.~(\ref{3SEqs}), and obtain
\begin{equation}
   \begin{array}{l}
      2i\lambda _1 \dot a_1 =-\lambda _1\Delta a_1+\lambda _2\Omega \sqrt{2}a_2,
      \\  i\lambda _2(\dot a_1a_2+a_1\dot a_2)
	     =\Omega \sqrt{2}(\lambda _1a_1^2+\lambda _3a_2^2), \\ 
      2i\lambda _3\dot a_2	
      =\lambda _2\Omega \sqrt{2}a_1+\lambda _3\Delta a_2.
   \end{array}
\end{equation}
By substituting $\dot a_1$ and $\dot a_2$, found from the first and third
equation, into the second we conclude that the latter will be satisfied
identically if $\lambda _2^2=2\lambda _1\lambda _3$. Furthermore, it is
readily seen that if we take $\lambda _1=\lambda _3=1,\lambda _2=\sqrt{2}$,
the first and third equations for $a_1$ and $a_2$ reduce exactly to
Eqs.~(\ref{2SEqs}). Thus the solution to the three-state equations~(\ref{3SEqs})
is indeed expressed in terms of the solution $(a_1,a_2)$ of the two-state
equations~(\ref{2SEqs}): $c_1=a_1^2,c_2=\sqrt{2}a_1a_2,c_3=a_2^2$.

The result for $N=3$ encourages us to try in the case of general $N$
the ansatz
\begin{equation}
   \label{ansatz-N}
   \begin{array}{l}
      c_1=\lambda _1a_1^{N-1} \\ 
      c_2=\lambda _2a_1^{N-2}a_2 \\ \ldots  \\ 
      c_n=\lambda _na_1^{N-n}a_2^{n-1} \\ \ldots  \\ 
      c_N=\lambda _Na_2^{N-1}
   \end{array}
\end{equation}
and we choose $\lambda _1=\lambda _N=1$. We substitute this ansatz in
Eq.~(\ref{SEqs}) and from the first and the last equations we find the 
following equations for $\dot a_1$ and $\dot a_2$
\begin{equation}
   \begin{array}{l}
      i(N-1)\dot a_1=-j\Delta a_1+\lambda _2\sqrt{N-1}\Omega a_2, \\ 
      i(N-1)\dot a_2=\lambda _{N-1}\sqrt{N-1}\Omega a_1+j\Delta a_2.
   \end{array}
\end{equation}

By substituting these derivatives in the equation for $\dot c_n$, we
conclude that the latter will be satisfied identically if
\begin{eqnarray}
   \label{Rel1}
   \lambda _{n-1}&=&\sqrt{\frac{n-1}{(N-1)(N-n+1)}}\lambda_{N-1}\lambda _n,\\
   \label{Rel2}
   \lambda _{n+1}&=&\sqrt{\frac{N-n}{n(N-1)}}\lambda _2\lambda_n.
\end{eqnarray}
By changing $n\rightarrow n+1$ in Eq.~(\ref{Rel1}) and multiplying it
with Eq.~(\ref{Rel2}) we find that 
\begin{equation}
   \label{Rel3}\lambda _2\lambda _{N-1}=N-1
\end{equation}
By applying Eq.~(\ref{Rel2}) repeatedly $n$ times, we obtain
\begin{equation}
   \label{Rel4}
   \lambda _{n+1}=
   \sqrt{\frac {(N-1)!}{n!(N-n-1)!} \frac 1{(N-1)^n}}\lambda _2^n,
\end{equation}
where we have accounted for $\lambda _1=1$. We now set $n=N-2$ in 
Eq.~(\ref{Rel4}), and taking Eq.~(\ref{Rel3}) into account we obtain 
$\lambda _2=\lambda_{N-1}=\sqrt{N-1}$. Then Eq.~(\ref{Rel4}) immediately gives
\begin{equation}
   \label{lambdas}
   \lambda _n=\sqrt{\frac {(N-1)!}{(n-1)!(N-n)!}},
\end{equation}
Thus, we conclude that the solution to the $N$-state equations~(\ref{SEqs})
is expressed in terms of the solution $(a_1,a_2)$ of the two-state
equations~(\ref{2SEqs}) by Eqs.~(\ref{ansatz-N}) with $\lambda _n$ given by
Eq.~(\ref {lambdas}). Furthermore, the $N$-state initial
conditions~(\ref{cond}) require the tw0-state initial conditions
$a_1(-\infty)=1, a_2(-\infty)=0$. This implies that the final populations 
$P_n=\left| c_n(+\infty)\right|^2$ are expressed in terms of the two-state 
transition probability $p=\left| a_2(+\infty)\right| ^2=1-\left| 
a_1(+\infty)\right|^2$ as
\begin{equation}
   \begin{array}{l}
      P_1=(1-p)^{N-1} \\ 
      P_2=(N-1)(1-p)^{N-2}p \\     \ldots  \\ 
      P_n=\frac {(N-1)!}{(n-1)!(N-n)!} (1-p)^{N-n}p^{n-1} \\ \ldots  \\ 
      P_N=p^{N-1}.
   \end{array}
   \label{finalP}
\end{equation}

\begin{figure}
\centerline{\psfig{width=80mm,file=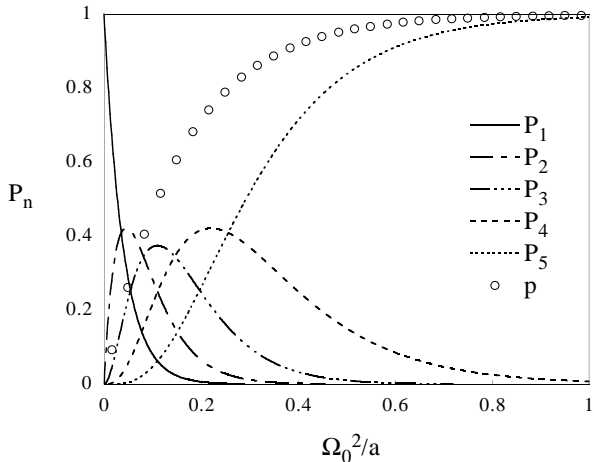}}
\caption[f3]{The transition probabilities for the $f=2$ multistate 
Landau-Zener model. The two-state solution is given by $p$, and goes clearly
faster to unity with increasing $\Omega_0^2/a$ than the corresponding 
five-state probability $P_5$. 
\label{kolme}}
\end{figure}

In estimating the spin-change probabilities for evaporative cooling we can use
the {\em Landau-Zener model}~\cite{LZ},
\begin{equation}
   \Delta (t)=at,\qquad \Omega (t)=\Omega_0=const, 
\end{equation}
where $a$ is proportional to the change in $\Delta(R)$ and to
the speed of atoms, both evaluated at the trap edge (at the rf
resonance). In the two-state model the transition probability is 
\begin{equation}
   p=1-\exp(-2\pi \Omega_0^2/a). 
\end{equation}
Thus our model provides the exact result for the transition
probabilities in the generalized multistate Landau-Zener model, which can be
used in estimating the efficiency of the evaporative cooling~\cite{evap}. As
an example we show the $f=2$ situation in Fig.~\ref{kolme}, where the final
populations are plotted as a function of the adiabaticity parameter
$\Omega_0^2/a$. As expected, in the multistate case one needs larger values
of $\Omega_0^2/a$ to achieve population inversion, than in the two-state case.

The Landau-Zener model can also describe the {\em chirped output coupling}.
Then the atoms are assumed to be stationary, so the time-dependence in $\Delta$
arises from the time-dependent change in $\omega_{\rm rf}$ (chirp), which
is typically linear in time. In Ref.~\cite{MITcoupler} the three-state version 
of the result~(\ref{finalP}) was successfully used in describing the 
corresponding experiment in sodium $f=1$ system. However, there the model was 
introduced only intuitively, and justified merely by a comparison with 
numerical solutions to Eq.~(\ref{SEqs}). Here we have provided the proof that
the $f=1$ solution is exact, and furthermore, derived the exact solution for
{\em any} $f$. The special case of the three-state Landau-Zener model is also
studied in Ref.~\cite{CH}.

Instead of a chirped field one can use {\em resonant pulsed output coupling},
which was also demonstrated in Ref.~\cite{MITcoupler}. For any resonant pulse 
we have $\Delta=0$ and thus the two-state system follows the area
theorem~\cite{Shore,GS}:
\begin{equation}
   \label{area}
   p = \sin^2(A/2),\qquad A= 2\int_{-\infty}^{\infty}dt\, \Omega(t),
\end{equation}
where $A$ is defined as the pulse area, typically $A\propto\Omega_0 T$ (here
$\Omega_0$ is the pulse peak amplitude and $T$ is the pulse duration). 
In the MIT experiment~\cite{MITcoupler} the number of atoms left in the trap 
oscillated as a function of the area of a resonant square pulse, exactly as
Eqs.~\ref{finalP} and~(\ref{area}) predict. However, our model is not limited
to resonant pulses only. For {\em off-resonant pulses} ($\Delta=const\neq 0$) 
in two-state systems there are several known analytic solutions, which are 
reviewed e.g.~in Refs.~\cite{Shore,GS}.

The purely time-dependent output coupler models described above are valid only 
if the time scales for the rf-induced interaction and the spatial dynamics of 
the condensate are very different. In molecular systems one expects interesting
effects when the excitation process and internal dynamics of the molecule 
couple~\cite{GS}. It might be possible to realize some of the 
predicted molecular wave packet phenomena using condensates. 

In this Letter we have shown that {\em any} time-dependent multistate model
describing the rf-induced coupling between the different atomic spins states
within the same hyperfine manifold can always be solved in terms of the
solution of the corresponding two-state model. The fact that these models play
a crucial role in time-dependent control of magnetically trapped ultracold 
atoms adds significantly to the importance of this result, which is
also in general quite fascinating.

This research has been supported by the Academy of Finland. K.-A. S. thanks Paul
Julienne for enlightening discussions on evaporative cooling.

\end{document}